\documentclass{aa}

\usepackage[varg]{txfonts}
\usepackage{amssymb}
\usepackage{upgreek}
\usepackage{graphicx}
\usepackage{epstopdf}
\usepackage{amsmath}
\usepackage{color}
\usepackage{natbib}
\usepackage{url}
\usepackage{hyperref}
\usepackage{gensymb}

\usepackage{cleveref}

\usepackage{bm}
\usepackage{mathtools}
\usepackage{lipsum}
\usepackage{physics}

\newcommand{\be}{\begin{equation}}
\newcommand{\ee}{\end{equation}}
\newcommand{\bc}{\begin{center}}
\newcommand{\ec}{\end{center}}
\newcommand{\beq}{\begin{eqnarray}}
\newcommand{\eeq}{\end{eqnarray}}

\bibpunct{(}{)}{;}{a}{}{,} 

\DeclareUnicodeCharacter{00A0}{ }

\makeatletter
\def\fvec#1{\underline{\sbox\tw@{$#1$}\dp\tw@\z@\box\tw@}}
\makeatother

\begin{document}

\title{Effects of scattering in the accretion funnel on the pulse profiles of accreting millisecond pulsars}

\titlerunning{Pulse profiles of accreting millisecond pulsars}

\author{Varpu Ahlberg\inst{1}
\and  Juri~Poutanen\inst{1} 
\and  Tuomo Salmi\inst{2}}

\authorrunning{V. Ahlberg et al.}

\institute{Department of Physics and Astronomy, FI-20014 University of Turku, Finland\\ \email{[varpu.a.ahlberg,juri.poutanen]@utu.fi} 
\and Anton Pannekoek Institute for Astronomy, University of Amsterdam, Science Park 904, 1098XH Amsterdam, the Netherlands
}
\date{Received 04 October 2023 / Accepted 16 November 2023}

\abstract{
The hotspot emission of accreting millisecond pulsars (AMPs) undergoes scattering in the accretion flow between the disk inner radius and the neutron star surface.
The scattering optical depth of the flow depends on the photon emission angle, which is a function of the pulse phase, and reaches its maximum when the hotspot is closest to the observer.
At sufficiently large optical depths the observed pulse profile should develop a secondary minimum, the depth of which depends on the accretion rate and the emission geometry.
Such a dip evolving with the accretion rate might explain the phase shift and pulse profile evolution observed in AMPs during outbursts.
Accounting for scattering is important for accurate modeling of the AMP pulse profiles in order to improve the accuracy of determination of the neutron star parameters, such as their masses and radii. 
In this paper we present a simplified analytical model for the Thomson optical depth of the accretion funnel, and apply it to simulating the pulse profiles.
We show that scattering in the accretion funnel has a significant effect on the pulse profiles at accretion rates of $\dot{M} \gtrsim 10^{-10}~{M}_\sun\, \mathrm{yr}^{-1}$.
Our model predicts a gradual evolution of the pulse profile with the accretion rate that appears to be consistent with the observations.
}


\maketitle

\section{Introduction}\label{sec:intro}

Accreting millisecond pulsars (AMPs) belong to a class of  neutron stars (NSs) in X-ray binaries rotating at millisecond periods and with a relatively low magnetic field of $10^8 - 10^9$~G.
They produce coherent pulsations from the hotspots that are created by magnetic funneling of the accretion flow \citep{Romanova2003,Kulkarni2005}.
AMP X-ray pulse profiles tend to be nearly sinusoidal with some rapid variations that are not well understood \citep[see][for a review]{Patruno2021}.
Specifically, during outbursts the pulse phase of some AMPs are known to shift by large amounts ($\sim0.2$ of the pulse periods) and the profile can deviate from a sine wave or even develop a double-peaked structure.
Previous works have proposed that the phase shifts might be caused by movements of the hotspot \citep{Hartman2008,Lamb2009} or the expansion of the inner disk radius making a secondary hotspot visible \citep{Poutanen2009,Ibragimov2009}.
Neither model can fully explain the pulse shape variations, and other phenomena must therefore be considered.
Changes in the disk--magnetosphere interaction \citep{Kajava2011} or magnetic octupoles becoming dominant \citep{Long2012} have been shown to replicate some of the observed behavior.
It has also been suggested that scattering of the hotspot emission in the accretion funnel might play a role as it can have a significant optical depth during periods of enhanced accretion \citep{Poutanen2003}.
The optical depth through the funnel should vary with rotational phase depending on the geometry of the accretion flow.

The mostly sinusoidal shape of AMP pulse profiles is consistent with a dipole magnetic field with a small misalignment from the rotational axis.
The magnetic field disrupts the accretion disk relatively close to the NS and channels the gas into the accretion funnels \citep[e.g.,][]{Ghosh1978,Romanova2004}.
The funnel scatters some of the hotspot emission depending on how much plasma the light passes through.
The largest scattering optical depth occurs when the hotspot normal is aligned with the direction to the observer and is consequently occluded by the accretion funnel.
The optical depth variation thus leads to periodic dimming of the flux, and might result in a secondary minimum at zero pulse phase (i.e., when the hotspot is closest to the observer).
The strength of this variation is flux dependent because the accretion funnel is denser at high accretion rates, and therefore the effect is most apparent during the peak of outbursts.
Its influence on the pulse evolution should be gradual unless a rapid change occurs in the accretion geometry.

To provide a specific example, this accretion funnel occlusion may have been observed during certain outbursts of AMP SAX J1808.4$-$3658 (hereafter J1808). 
Its pulse profile changes significantly at certain outburst stages and sometimes exhibits a secondary minimum near the outburst peak \citep{Hartman2008,Ibragimov2009}.
This minimum was observed during the 2002 outburst of J1808 in the 10--24 keV band by \citealt{Ibragimov2009}, who noted that it was unlikely to be a result of a secondary hotspot as the accretion rate was high and the 2--3.7 keV profiles remained sinusoidal.
The absence of the secondary minimum in the soft X-ray band suggests that the funnel may not occlude the entirety of the hotspot, for example because the softer radiation is emitted by a larger area at the NS surface \citep{Gierlinski2005}. 
 The nonsinusoidal pulse profile during the early stages of the 2008 outburst of J1808 also casts doubt on a secondary hotspot being responsible as it did not remain nonsinusoidal at lower fluxes \citep{Hartman2009b}.
Variations in the disk--magnetosphere interaction have been shown to be a likely factor in the rapid pulse amplitude shift during this outburst \citep{Kajava2011}.
While previous works have considered funnel scattering as a likely factor, the lack of a quantitative model means that its impact is unknown.

A model for the effects of scattering would have applications for obtaining AMP parameter constraints via pulse profile modeling.
Importantly, measuring the masses and radii of NSs can be used to constrain the equation of state (EoS) of cold dense  matter \citep{Poutanen2003,Leahy2008,Leahy2011,Watts2016,Watts2019}.
Being able to model pulse profiles of AMPs showing thermonuclear X-ray bursts would be especially valuable as X-ray bursts can be used to measure the mass and radius using a different method based on the spectral evolution during the cooling tail \citep{Suleimanov2011,Suleimanov2017,Suleimanov2020,Nattila2017}.
Combining the methods could potentially increase the accuracy of mass--radius measurement and lead to more stringent constraints on the EoS.

In this paper we present a simplified model for calculating the scattering optical depth of AMP accretion funnels.
We apply it to simulated pulse profiles and study how scattering affects the AMP pulse shape at different accretion rates.
We show that the scattering may explain some of the observed features of the AMP pulse profile evolution.

\section{Scattering in the accretion flow}

\subsection{Optical depth} \label{sec:opticaldepth}

The optical depth $\tau$ of the accretion funnel depends on the light trajectory, the funnel geometry, and the properties of the accreted gas.
It is defined as
\begin{equation}
    \,\mathrm{d} \tau = \kappa \rho \,\mathrm{d}s,
\end{equation}
where $\kappa$ is the opacity, $\rho$ is the density, and $\mathrm{d}s$ is the path element.
Because the gas is moving at relativistic velocities, it is important to transform the opacity from the gas comoving frame to the laboratory frame. This can be done with
\begin{align}
    \kappa &= (1 - \mu \beta) \kappa',\\
    \beta &= v_\mathrm{ff}/c, \\
    v_\mathrm{ff}(r) &= \sqrt{\dfrac{2GM}{r}},
\end{align}
where $\kappa'$ is the opacity in the comoving frame, $v_\mathrm{ff}$ the plasma free-fall velocity, $r$ the radial coordinate from the star center, and $\mu$ the cosine angle between the light trajectory and the gas velocity.  
We note that in AMPs a typical area of the hotspot is about two orders of magnitude larger \citep{Ibragimov2009, Lamb2009,Kulkarni2013} than in classical strong magnetic field pulsars,  resulting in a negligible effect of radiation pressure on the gas dynamics \citep{Lyubarskiy82,Mushtukov15}. 
This justifies our assumption of the free-fall velocity.
Close to the NS surface, at sufficiently high luminosities, the flow should attain the Compton temperature that is defined by the balance of Compton heating and cooling.  
Under these conditions, the accretion flow is fully ionized and the opacity is dominated by electron scattering. 
For the purposes of this paper, we use the Thomson scattering opacity for a pure hydrogen atmosphere $\kappa'=\kappa_\mathrm{e} = 0.4$~cm$^2$\,g$^{-1}$ as the choice does not change the qualitative behavior of our model.

Under these simplifications the only factors that need further consideration are the density of the gas and the light trajectory.
The density in the laboratory frame can be approximated using the continuity equation.
Assuming two identical funnels above two hotspots, it is
\begin{equation}
     \rho(r) = \frac{\dot{M}}{2 A(r) v_\mathrm{ff}(r) } ,
\end{equation}
where $A(r)$ is the area of one of the accretion funnels as a function of radius from the center of the star, the factor of 2 representing the total area of the two funnels.

The light travel time through the funnel is small enough compared to the rotation of the hotspot that we approximate it as instantaneous.
In the Schwarzschild metric in the equatorial plane and at constant time, the path element is $\mathrm{d}s^2 = \mathrm{d}r^2 (1-R_\mathrm{S}/r)^{-1} + r^2 \mathrm{d} \psi^2$, where $R_\mathrm{S} = 2GM/c^2$ is the Schwarzschild radius.
Using the azimuth coordinate $\psi$ as a parameter, the optical depth integral can be written as 
\begin{equation} \label{eq:tauintegr}
    \tau = \kappa_\mathrm{e} \int_{\psi_\mathrm{min}}^{\psi_\mathrm{max}} \rho (1 - \mu \beta) \,\mathrm{d} \psi \left[\left(\dfrac{\mathrm{d}r}{\mathrm{d} \psi}\right)^2 (1 - R_\mathrm{S}/r)^{-1} + r^2 \right]^{1/2},
\end{equation}
where $\psi_\mathrm{max}$ and $\psi_\mathrm{min}$ are the range of the trajectory. 
The derivative of the radius term is known analytically (\citealt{Pechenick83}; \citealt{Misner73}, p. 674)
\begin{equation} 
\dfrac{\mathrm{d}r}{\mathrm{d}\psi} = \frac{r^2}{b} \left[  1 - b^2 (1-R_S/r)/r^2
\right]^{1/2} ,
\end{equation} 
where $b$ is the impact parameter. 
Thus, the optical depth is 
\begin{equation} \label{eq:tauintegr2}
    \tau = \kappa_\mathrm{e} \int_{\psi_\mathrm{min}}^{\psi_\mathrm{max}}  \rho (1 - \mu \beta) \,   \frac{r^2\,\mathrm{d} \psi}{b \sqrt{1-R_\mathrm{S}/r}} .
\end{equation} 
We calculate the intensity of light using a simplified equation for the radiative transfer $I = I_0 {\rm e}^{-\tau}$, where $I_0$ is the intensity of the initial emission and $I$ the intensity of light that passes through unscattered.
This approximation does not account for the light scattered into the direction of the observer from other directions and should be interpreted as an estimate for the dimming of the light rather than an accurate model for electron scattering.

\subsection{Light trajectory}

\begin{figure}
    \centering
    \includegraphics{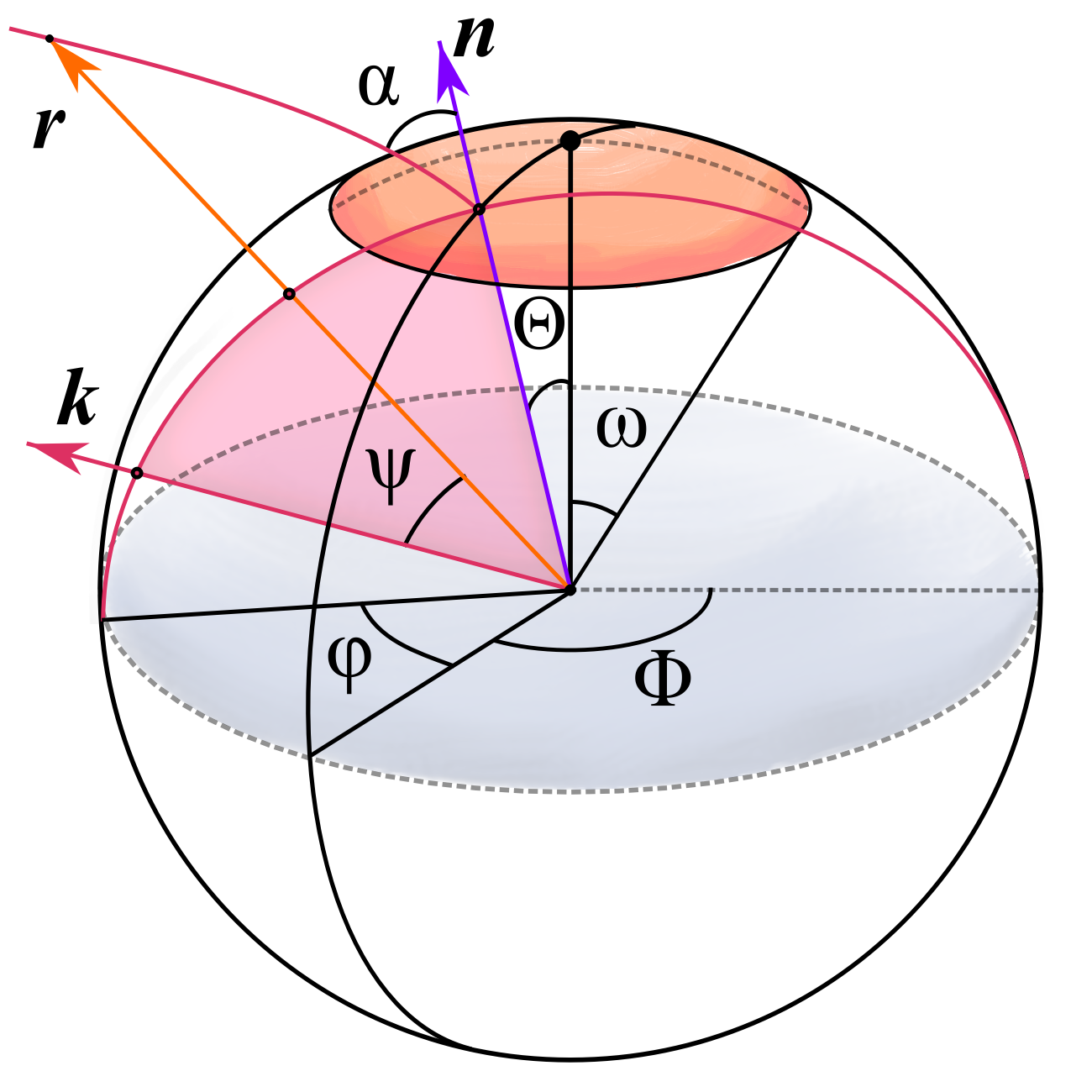}
    \caption{Geometry of the hotspot and the light trajectory in spherical coordinates aligned with the spot center. The plane of emission and the trajectory are drawn in magenta, and the circular hotspot is in orange.}
    \label{fig:geometry}
\end{figure}

Calculating the integral limits in Eq. \ref{eq:tauintegr2} requires some parameterization of the light trajectory in order to solve the point where it exits the funnel.
This can be achieved using a vector model.
We define unit vectors for the direction of a distant observer, $\vec{\hat k}$, and for the surface normal of the emitting point, $\vec{\hat n}$.
The angle between the normal and the emitted photon at infinity is $\cos \psi_\infty = \vec{\hat k} \cdot \vec{\hat n}$, which in this case is the limit $\psi_\mathrm{min}$. 
We let $\vec{\hat r}$ be a vector pointing to some point along the trajectory with a known angle $\cos \psi = \vec{\hat k} \cdot \vec{\hat r}$.
In the Schwarzschild metric the light trajectory lies on the plane defined by $\vec{\hat k}$ and $\vec{\hat n}$, so $\vec{\hat r}$ can be calculated as a linear combination of the two vectors:
\begin{equation} \label{eq:lpsi}
    \vec{\hat r} = \dfrac{ \vec{\hat n}\sin \psi  + \vec{\hat k} \sin (\psi_\infty - \psi) }{\sin \psi_\infty}.
\end{equation}
To calculate these vectors we use a coordinate system where the center of the hotspot is aligned with the z-axis.
The geometry in the spot center coordinates is shown in Fig.~\ref{fig:geometry}.
The normal vector can now be written as $\vec{\hat n} = (\sin \Theta \cos \Phi, \sin \Theta \sin \Phi, \cos \Theta)$, where $\Theta$ and $\Phi$ are the colatitude and azimuth of the emitting element.
To define $\vec{\hat k}$, we use $\psi_\infty$ and $\varphi$ as the inclination and azimuth of the observer relative to $\vec{\hat n}$.
This vector can be found using the following rotation matrix
\begin{align}
    \vec{\hat k} &= \mathbf{R}(\Phi, \Theta) \cdot \begin{pmatrix}
        \sin \psi_\infty  \cos \varphi \\
        \sin \psi_\infty\sin \varphi  \\
        \cos \psi_\infty
    \end{pmatrix}, \\
    \mathbf{R}(\Phi, \Theta) &= \begin{pmatrix}
    \cos \Theta \cos \Phi & -\sin \Phi & \sin \Theta \cos \Phi\\
    \cos \Theta \sin \Phi & \cos \Phi & \sin \Theta \sin \Phi\\
    -\sin \Theta & 0 & \cos \Theta
    \end{pmatrix}. \label{eq:mrot}
\end{align}

\noindent These radial vectors are accurate in the Schwarzschild metric as spherically symmetric spacetimes are isomorphic to 3D rotations.
Assuming a circular hotspot, the geometry will be symmetric around $\Phi$ so only $\Phi = 0$ has to be considered.
Thus, we can write
\begin{align}
    \vec{\hat k} &= \begin{pmatrix}
        \cos \psi_\infty \sin \Theta + \cos \varphi \sin \psi_\infty \cos \Theta \\
        \sin \varphi \sin \psi_\infty \\
        \cos \psi_\infty \cos \Theta - \cos \varphi \sin \psi_\infty \sin \Theta
    \end{pmatrix}.
\end{align}
It follows that
\begin{equation} \label{eq:trajectory}
    \vec{\hat r} = \begin{pmatrix}
        \cos (\psi_\infty - \psi) \sin \Theta + \sin (\psi_\infty - \psi) \cos \Theta \cos \varphi \\
        \sin (\psi_\infty - \psi) \sin \varphi \\ 
        \cos (\psi_\infty - \psi) \cos \Theta - \sin (\psi_\infty - \psi) \sin \Theta \cos \varphi
    \end{pmatrix}.
\end{equation}
The angle between $\vec{\hat r}$ and the hotspot center $\cos \theta = \vec{\hat r} \cdot (0,0,1)$ can now be expressed as a function of $\psi$:
\begin{align} \label{eq:thetapsi}
    \cos \theta &= A \cos \psi + B \sin \psi, \\
    A &= \cos \psi_\infty \cos \Theta - \sin \psi_\infty \sin \Theta \cos \varphi, \\
    B &= \sin \psi_\infty \cos \Theta + \cos \psi_\infty \sin \Theta \cos \varphi.
\end{align}
Reversing this equation yields $\psi$ as a function of $\theta$, which is helpful in finding the exit point $\psi_\mathrm{max}$ of the trajectory:
\begin{equation} \label{eq:psitheta}
    \tan \dfrac{\psi}{2} = \dfrac{B \pm \sqrt{\cos^2 \Theta + \sin^2 \Theta \cos^2 \varphi - \cos^2 \theta}}{A + \cos \theta}.
\end{equation}The radial coordinate of the trajectory as a function of $\psi$ in a flat space time can be solved with basic trigonometry,
\begin{equation} \label{eq:linear}
    r(\psi) = \frac{R \sin \alpha}{\sin\psi},
\end{equation}
where $\alpha$ is the angle between $\vec{\hat n}$ and the initial direction of emission $\vec{\hat k}_0$, which in this case is the same as $\psi_\infty$.
In the Schwarzschild metric, one approximation for the trajectory including light bending is \citep{Beloborodov_2002}
\begin{equation} \label{rpsi}
    r(\psi) = \left[ \frac{R_\mathrm{S}^2 (1-\cos \psi)^2}{4(1 + \cos \psi)^2} + \frac{b^2}{\sin^2 \psi} \right]^{1/2} - \frac{R_\mathrm{S} (1-\cos \psi)}{2(1+ \cos \psi)} , 
\end{equation}
where the impact parameter is 
\begin{equation} \label{eq:impact}
b = \frac{R}{\sqrt{1-R_\mathrm{S}/R}} \sin \alpha ,
\end{equation}
\begin{equation} \label{eq:beloborodov}
1 - \cos \alpha = x = (1-\cos \psi) (1- R_\mathrm{S}/r) \equiv y(1-u),
\end{equation} 
where $u = R_\mathrm{S}/r$ and $y = 1 - \cos \psi$.
This formula is fairly accurate for low $\alpha$ angles as long as $r>2R_\mathrm{S}$.
The light bending formula described in \cite{Poutanen2020} is more accurate:
\begin{equation} \label{poutanen}
    x = (1-u) y \left( 1 + \dfrac{u^2 y^2}{112} - \dfrac{\rm e}{100}uy \left[ \ln \left( 1 - \dfrac{y}{2} \right) + \dfrac{y}{2} \right] \right).
\end{equation}
Here \text{e} is  Euler's number.
The increased mathematical complexity makes it impossible to solve $r(\psi)$ analytically, and we therefore use Eq.~\eqref{rpsi} to evaluate the integrand in Eq.~\eqref{eq:tauintegr2}.
This integral must be done numerically, but it is smooth and contains no singularities within integral limits.
The impact parameter can be calculated from Eq.~\eqref{eq:impact}, and the angle $\alpha$ can be obtained  from either Eq.~\eqref{poutanen} or Eq.~\eqref{eq:beloborodov}.

\subsection{Accretion funnel and hotspot geometry}

Three-dimensional numerical simulations of the accretion flow show different hotspot shapes depending on the magnetic obliquity $\theta_0$ \citep{Romanova2003, Romanova2004}. 
For the obliquity of $\theta_0 \approx 30\degr$, the shape of the hotspot is a hollow semicircle centered around the magnetic pole.
At lower obliquities, the hotspot becomes a hollow circle, and at very high angles it starts to become bar-shaped.
The angle between the magnetic pole and the center of the hotspot for a dipole field is $\theta_\mathrm{s} = \arcsin\left( \cos{\theta_0} \sqrt{{R}/{R_{\rm M}}}  \right)$, where $R_{\rm M}$ is the magnetospheric radius.
The actual magnetic field lines deviate from a perfect dipole as they are compressed by the disk--magnetosphere interaction.
In simulations the width of the spot is about $7\degr-8\degr$ and is only weakly dependent on other parameters \citep{Kulkarni2013}.
This geometry would add several free parameters to our model to account for the hotspot shape, and  so we do not consider it here.

A difficulty with modeling the shape of the accretion funnel's edge is its motion relative to the observer and the emitting spot.
If the geometry is defined in a frame comoving with the hotspot center, solving it in the lab frame requires a Lorentz transform.
As the true shape of the funnel is unknown, we define all geometry here in the lab frame.
While this ignores the effects of rotation, it does not impose a greater error than the simplified geometry already does.

\begin{figure*}
    \centering
    \includegraphics[width=0.8\textwidth]{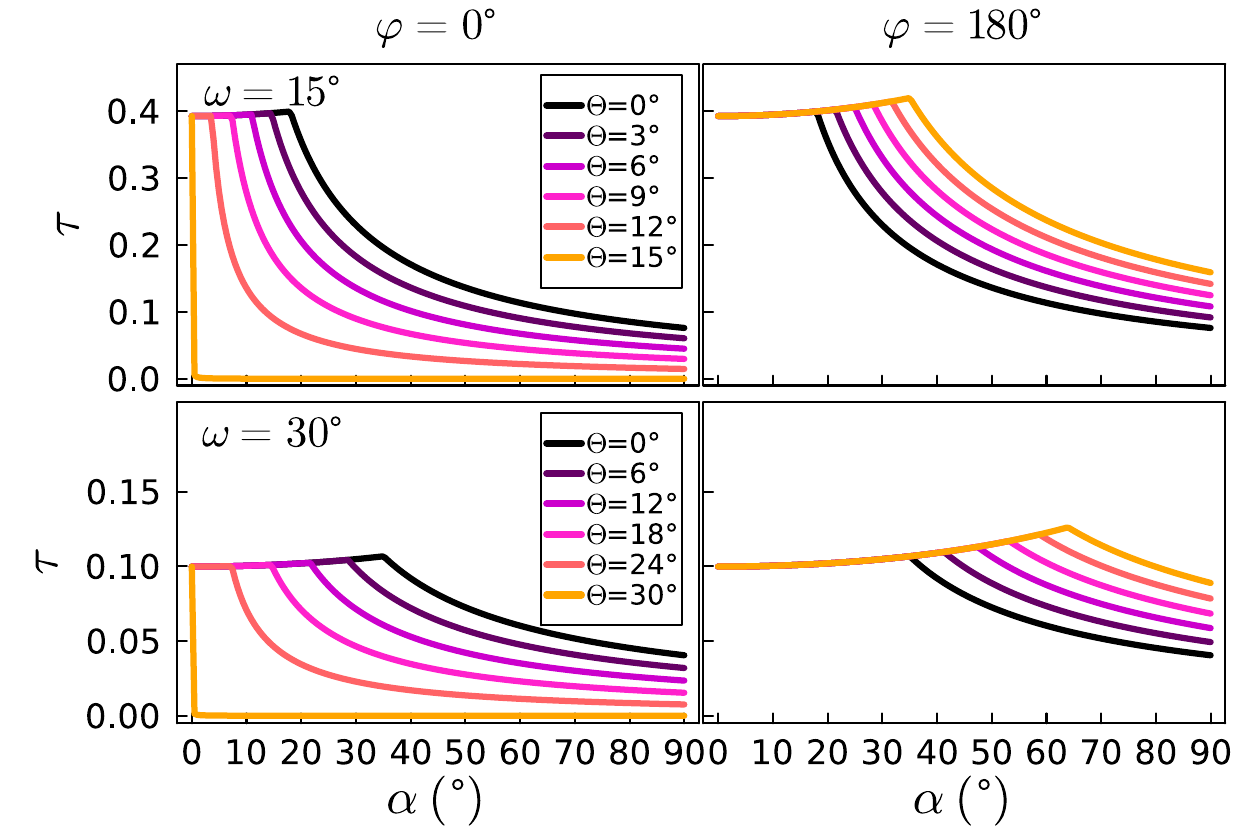}
    \caption{Optical depth of a conical accretion funnel as a function of emission angle $\alpha$ at azimuth angles  $\varphi=0\degr$ (left; photon is emitted away from the spot center) and $180\degr$ (right; photon is emitted toward the spot center) and several angular separations $\Theta$ (different colors) for spot sizes of $\omega=15\degr$ (top) and $30\degr$ (bottom) at  $\dot{M}_{10}=1$. 
    }
    \label{fig:tables}
\end{figure*}

Approximating the accretion funnel edge by magnetic field lines also has its challenges.
In a classical perfect dipole the field lines follow $r = R_\mathrm{M} \sin^2 \theta$. 
This leads to a funnel area of $A(r) = 2 \pi r^2 (1 - \sqrt{1-r/R_\mathrm{M}})$.
However, the classical dipole does not take the curvature of the metric into account, and the cosine angle between the light trajectory and the gas velocity $\mu$ is difficult to solve.
The light trajectory can cross the curved funnel border several times, and the exit points are not analytically solvable.
Field lines originating from a circular hot spot cannot set a clear upper boundary for the funnel, and so a maximum radius must be chosen.
It should be smaller than the magnetospheric radius, which is on the order of a few $R$ for AMPs.

To simplify further, we model the funnel as a circular cone with a half-opening angle of $\omega$ and an area of $A = 2\pi r^2(1-\cos \omega)$.
The exit point of the trajectory is easily solvable with Eq. \eqref{eq:psitheta} by setting $\theta = \omega$.
We model the gas velocity to be in the radial direction, and thus the angle between the light trajectory and the velocity is $\mu = -\cos (\alpha(\psi, r))$, which has several analytical approximations.
We use the more accurate formula \eqref{poutanen} to calculate this angle, but the radius has to be solved using Eq.~\eqref{rpsi}. 
The error coming from the approximate radius is negligible compared to the error of calculating the angle using the less accurate formula.
Setting the upper radius of the cone affects the maximum optical depth, but it tends to converge at around $r_\mathrm{max} = 5R$.
We consider a limiting radius of $r_\mathrm{max}=3R$ for the purposes of this paper as it better represents AMP inner disk radii.
In the dipole model $A \propto r^3$, whereas in the conical model $A \propto r^2$, which makes it a higher limit for the optical depth compared to the dipole model.
As the actual field lines are less curved close to the pole than in a dipole, the accurate area of the funnel near the surface is somewhere between the perfect dipole and the cone.

\subsection{Pulse profile model}

To generate pulse profiles that incorporate our scattering optical depth model, we adapted the methodology described in \citet{Salmi2021}. 
The energy dependent observed flux of the hotspot from an emitting area $dS'$ is \citep{Poutanen2006}
\begin{equation}
    \mathrm{d} F_{E} = (1-u)^{1/2} \delta^4 I'_{E'}(\alpha') \cos \alpha \dfrac{\mathrm{d} \cos \alpha~~}{\mathrm{d} \cos \psi_\infty} \dfrac{\mathrm{d} S'}{D^2},
\end{equation}
where $u = R_\mathrm{S}/R$, $D$ is the distance of the observer, and $I'_{E'} (\alpha')$ the emitted intensity as a function of emission angle in the corotating frame.
The Doppler factor $\delta$ is defined as
\begin{equation}
    \delta = \dfrac{1}{\gamma(1-\beta_\mathrm{s} \cos \xi)},
\end{equation}
where $\beta_\mathrm{s} = v/c$ is the surface velocity, $\xi$ is the angle between the direction of emission and the velocity, and $\gamma = (1-\beta_\mathrm{s}^2)^{-1/2}$ is the Lorentz factor.
Pulse profiles are commonly modeled using the oblate Schwarzschild approximation, that is, calculating the light bending and the corresponding time delays accurately for a Schwarzschild metric and correcting for the star's oblateness (see \citealt{Loktev2020} and \citealt{AlGendy2014}).
We follow a similar method, but we neglect oblateness to be consistent with the spherical geometry assumed in our simplified scattering model.

We modeled the hotspot emission as a  slab of hot   (50~keV)  electrons of Thomson optical depth 1, lying above a blackbody surface of temperature of 1 keV as determined from the observed X-ray spectra \citep[e.g.,][]{Gierlinski2005,Ibragimov2009}.
The resulting spectrum is a combination of unscattered blackbody photons and a Comptonized component at high energies.
We determined the energy and angular dependence of radiation using an approximate Thomson scattering model, and prescribing the energy increase in each Compton scattering according to a standard formula \citep{Viironen2004,Salmi2021}.
We interpolate the results presented in the tables of \citet{Salmi2021}.
The dependence of the seed photon intensity on the emission angle $\alpha'$ in the comoving frame  was modeled using an approximation for the semi-infinite electron scattering dominated atmosphere: $I'(\alpha') \propto 0.421 + 0.868 \cos\alpha'$ \citep{Chandrasekhar1947}.
Choosing this over an isotropic distribution has a minor effect on the pulse profile, but it should not matter for the purposes of this paper.

\begin{figure*}
    \centering
    \includegraphics[width=0.8\textwidth]{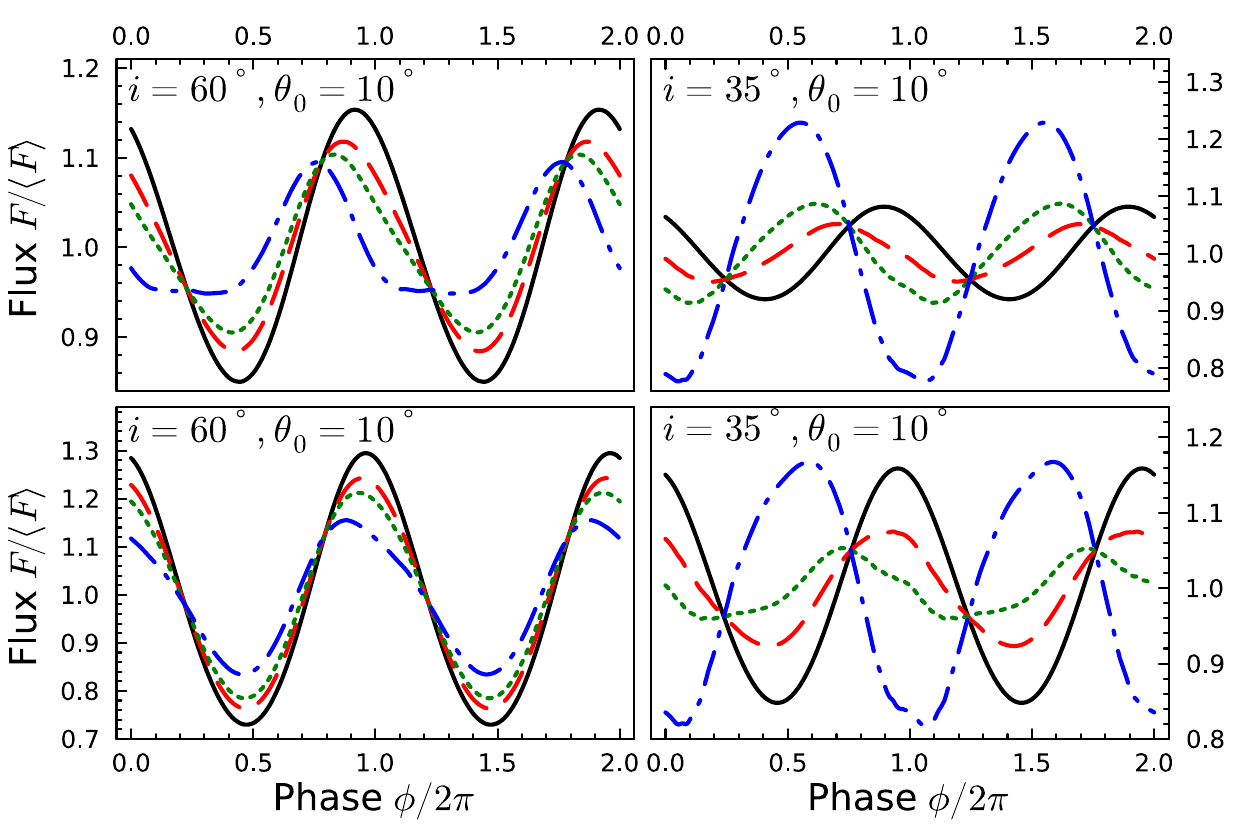}
    \caption{Simulated pulse profiles for different geometrical parameters and accretion rates using one hotspot with an angular radius of $\omega = 25\degr$ at an energy of 10 keV (top) and 3 keV (bottom). The solid, dashed, dotted, and dash-dotted lines (black, red, green, and blue, respectively)  correspond to accretion rates of $\dot{M}_{10}=0$, $3$, $5$, and $10$, respectively.}
    \label{fig:profiles1}
\end{figure*}

We calculated the optical depth integral individually for each emitting element of the hotspot at each phase point.
In order to find the range of the trajectory, the vectors $\vec{\hat n}$ and $\vec{\hat k}$ must be calculated in the hotspot center coordinates.
The pulse profile model defines them in coordinates where the observer $\vec{\hat k}^*=(\sin i, 0, \cos i)$ is fixed at an inclination of $i$ and the normal vector $\vec{\hat n}^*$ rotates with the star.
The observer direction can be obtained with the   rotation
\begin{equation}
    \vec{\hat k} = \mathbf{R}(\phi, \theta_0)^\mathrm{T} \cdot \vec{\hat k}^*,
\end{equation}
where $\theta_0$ is the magnetic obliquity of the spot center and $\phi$ is the phase angle.
Using Eq.~\eqref{eq:lpsi} and defining $\vec{\hat n}$   as before, we find
\begin{align}
    \cos \theta &= A \cos \psi + B \sin \psi , \\
    A &= \sin i \sin \theta_0 \cos \phi + \cos i \cos \theta_0 , \\
    B &= \frac{\cos \Theta - \cos \psi_\infty A}{\sin \psi_\infty}.
\end{align}
Similarly to Eq.~\eqref{eq:psitheta}, reversing this leads to
\begin{align}
    \tan \dfrac{\psi}{2} &= \dfrac{B \pm \sqrt{A^2 + B^2 - \cos^2 \theta}}{A + \cos \theta}.
\end{align}
Thus, we can find the integral range with $i$, $\phi$, and $\theta_0$ instead of $\varphi$.
Finally, we scaled the emitted intensities by a factor of $e^{-\tau}$, as explained in Sect. \ref{sec:opticaldepth}, taking time delays into account.

\section{Applications}

\subsection{Dependence of optical depth on model parameters}

We tested the angular dependence of the scattering optical depth with different hotspot sizes.
We fixed the accretion rate at $\dot{M} = 10^{-10}~ {M}_\sun \, \mathrm{yr}^{-1}$;  it can be easily changed by scaling the optical depth with a factor of $\dot{M}_{10}$ (accretion rate in units of $10^{-10}~ {M}_\sun\, \mathrm{yr}^{-1}$). The mass and radius of the NS were set at $M = 1.4\,{M}_\sun$ and $R = 12\,~\mathrm{km}$.
The plots for the optical depth as a function of $\alpha$ at fixed $\varphi$ at spot radii $\omega$ of $30\degr$ and $15\degr$  are presented in Fig.~\ref{fig:tables}. 
The respective range of optical depths for the two spot sizes are about 0.02--0.1 and 0.05--0.4.
The maximum optical depth can be approximated as
\begin{equation}
    \tau_\mathrm{max} \approx 0.22 \left( \frac{20^\circ}{\omega}\right)^2 \dot{M}_{10},
\end{equation}
where the factor of 0.22 is related to the vertical optical depth from the spot center to the funnel upper boundary, and the parameter dependence is estimated from the gas density.
Therefore, if the accretion rate $\dot{M}_{10}$ is significantly lower than 1, the optical depth will be negligible according to our model.
If the accretion rate is much higher than 10, the funnel will no longer be optically thin at low emission angles for moderately sized spots.
The simplified scattering model is no longer valid when $\tau > 1$, and thus it is not physically accurate past $\dot{M}_{10} \sim 10$.

The optical depth increases at lower $\alpha$ as a result of the light passing through more of the funnel, reaching a plateau as it starts exiting through the upper edge.
The width of this plateau depends on the range of angles when the upper edge is reached, which is determined by $\omega$, $r_\mathrm{max}$, $\Theta$, and $\varphi$.
The plateau is not entirely flat since the distance the light travels through the gas is shorter at low emission angles.
The dependence of the optical depth on $\varphi$ increases toward the edges of the spot since it has a stronger influence on the light travel distance.
The effect that the spot size has on pulse profiles has previously been argued to be weak for spot radii of $\omega \lesssim 45\degr$ \citep{Lamb2009}, and yet in this case it has a major impact on the optical depth.
For small spots the density of the funnel is related to the spot size by approximately $\rho \propto \omega^{-2}$, and leads to overall larger optical depths despite shorter travel distances.
As accretion rate and spot size have a similar effect on the optical depth it may be difficult to separate the parameters in model fits.

\begin{figure}
\centering
\includegraphics[width=0.8\linewidth]{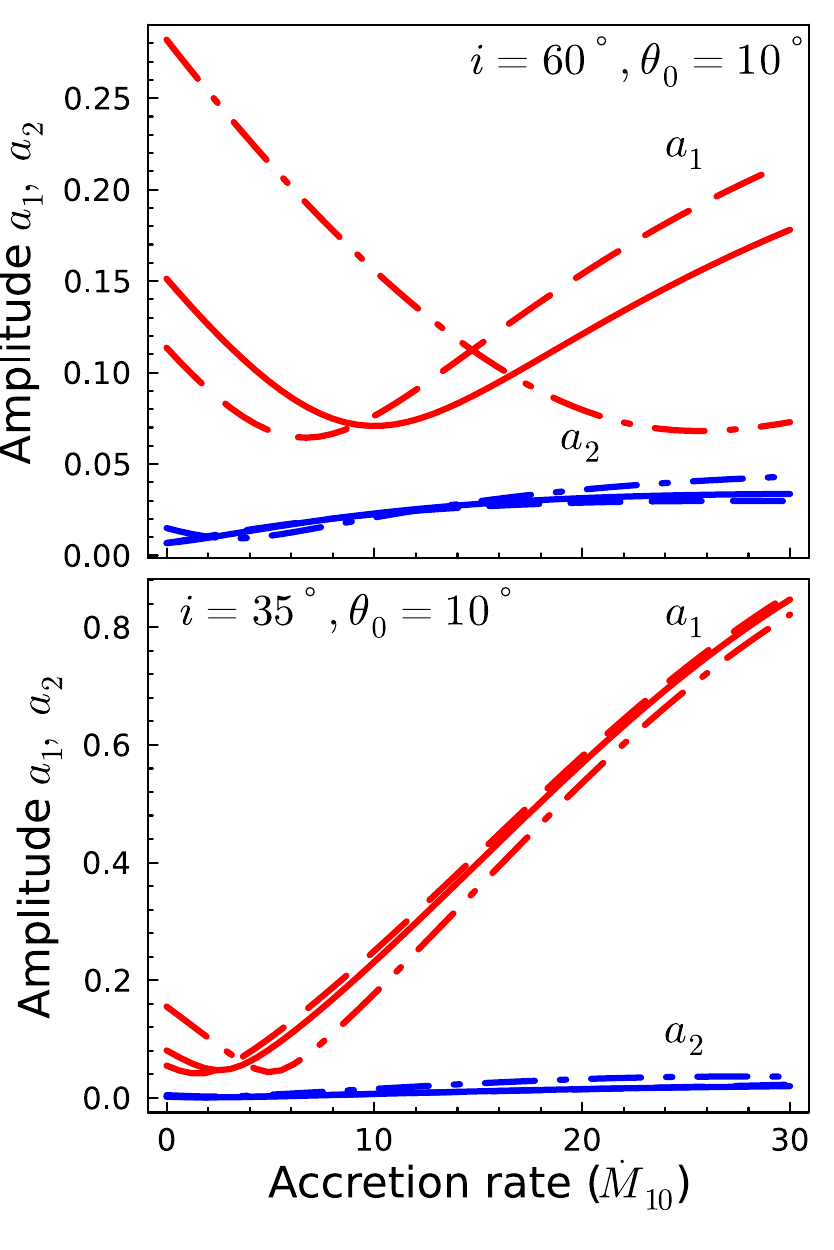}
    \caption{Amplitudes $a_1$ (red) and $a_2$ (blue) as a function of accretion rate for $i = 60\degr$  (top) and $i = 35\degr$ (bottom) for a single hotspot with an angular radius of $\omega = 25\degr$ and magnetic obliquity of $\theta_0 = 10\degr$. The dashed, solid, and dash-dotted lines correspond to energies of 20, 10, and 3 keV, respectively. 
    }
    \label{fig:amplitudes}
\end{figure}

\begin{figure}
\centering
\includegraphics[width=0.8\linewidth]{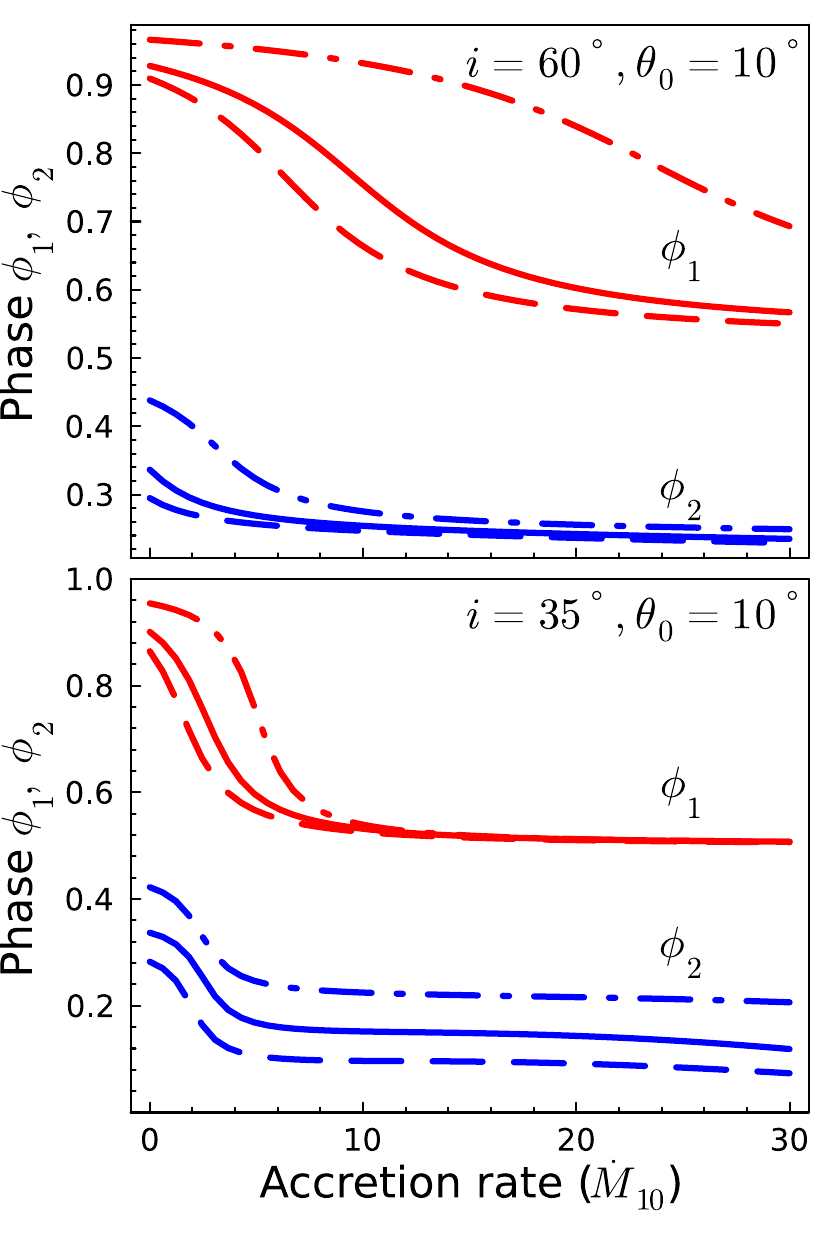}
    \caption{Same as Fig. \ref{fig:amplitudes}, but for the phase shifts $\phi_1$ (red) and $\phi_2$ (blue). }
    \label{fig:shifts}
\end{figure}

\subsection{Pulse profiles}

We generated various pulse profiles with the optical depth correction using different parameters.
We fixed the spin rate of the NS at $401\,\mathrm{Hz}$ (corresponding to  J1808) and the mass and radius at $1.4~{M}_\sun$ and $12~\mathrm{km}$, respectively.
Figure \ref{fig:profiles1} shows example pulse profiles with different parameters at photon energies of $10~\mathrm{keV}$ and $3~\mathrm{keV}$.
As predicted, the scattering in the accretion funnel   indeed causes a periodic decrease in flux at zero phase (hereafter called a dip) if the accretion rate is high.
The center of the dip remains at the same phase, while the pulse maximum arrives earlier at higher energies, causing the dip to appear after the peak.
The overall strength of the dip also depends on the amplitude of the pulse and may even dominate the rest of the profile if the amplitude is comparably low.
Interestingly, the dip seems to push the pulse maximum to earlier phases at higher accretion rates.
We investigate the effect the accretion rate has on the phase shifts and the energy dependence of the profile in Sect.~\ref{sec:phase}. 

Large differences in $i$ and $\theta_0$ lead to broad dips that do not cause clear secondary minima.
Conversely, when the angles are close to one another the dip becomes more prominent. 
The dip has an especially sharp shape when $|i-\theta_0| < \omega$, as some of the light exits through the upper edge of the funnel.
Because  the optical depth at the upper edge is smaller at low $\alpha$, the minimum of the dip occurs slightly before or after zero phase if the light passes the edge.
This behavior is a result of the simplified funnel and hotspot geometry used by our model and is not indicative of the actual physical scenario.
It still shows that clear secondary minima will most likely occur when the observer inclination is close to the magnetic obliquity.

The scattering optical depth has little effect on the pulse profile shapes of secondary hot spots as they are observed at shallow angles.
They may exhibit secondary minima if both the observer and the hotspot are nearly perpendicular to the axis of rotation, which is an unlikely scenario for AMPs.
Even though their scattering optical depth is only weakly phase dependent, it should not be neglected as it influences the amplitude of the pulse.

\subsection{Pulse amplitude and phase dependence on accretion rate} \label{sec:phase}

To test pulse phase characteristics as a function of accretion rate we fitted a cosine profile to the simulated pulse profiles
\begin{equation}
    F(\phi) = \langle F \rangle \{ 1 + a_1 \cos [2 \pi (\phi - \phi_1)] + a_2 \cos [4 \pi (\phi - \phi_2)] \}, 
\end{equation}
where $a_1$ and $a_2$ are the amplitudes of the fundamental and the first harmonic, $\phi_1$ and $\phi_2$ their phase shifts, and $\langle F \rangle$ the average flux.
Figures \ref{fig:amplitudes} and \ref{fig:shifts} show the amplitudes and phases over a wide range of accretion rates in different energy bands.
The amplitude of the fundamental first decreases, then eventually increases as the dip begins dominating the pulse characteristics.
The phases follow a similar pattern, decreasing smoothly down to a constant value at high accretion rates.
At accretion rates of $\dot{M}_{10}=0-5$ and at an inclination of $60\degree$ the phase of the fundamental barely changes in the low energy  band and shifts by $\sim 0.1$ in the higher bands.
This is linked with the amplitude of the fundamental beginning its increase at much higher accretion rates at low energies.
The amplitude of the harmonic is only weakly dependent on energy and increases slightly with accretion rate.
Its phase flattens out earlier than that of the fundamental and shifts more steeply in the low energy band.

\begin{figure}
    \centering
    \includegraphics[width=0.8\linewidth]{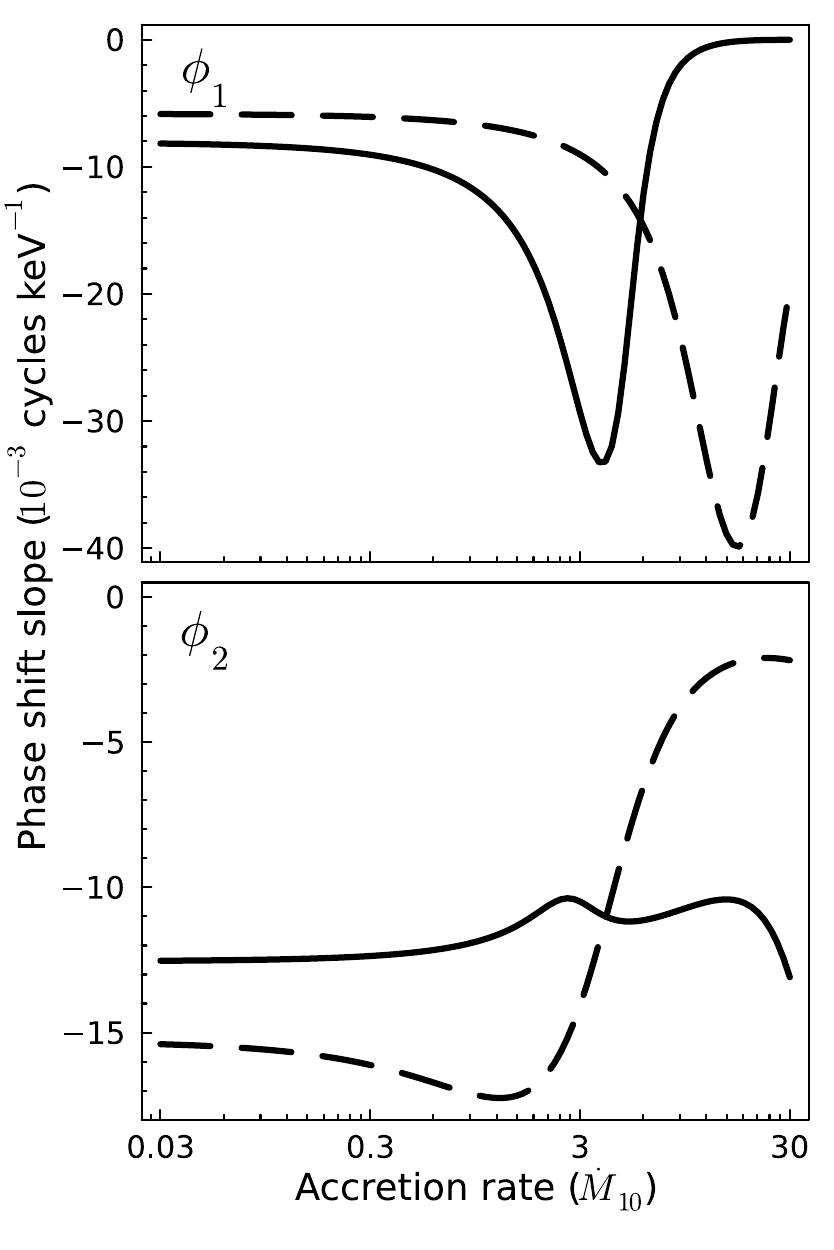}
    \caption{Plot of the slope of the phase shift of the fundamental (top) and harmonic (bottom) in the range 3--10 keV as a function of the accretion rate with geometrical parameters of $\theta_0 = 10\degr$ and $\omega=25\degr$, and at two different inclinations $i = 35\degr$ (solid) and $60\degr$ (dashed).
    }
    \label{fig:slopes}
\end{figure}

Interestingly, the phase lag between different energy bins changes with the accretion rate.
To investigate this behavior, we made a linear fit to the lags relative to the soft 3 keV band between energies of 3 and 10 keV.
The slope of the phase lags as a function of accretion rate is shown in Fig.~\ref{fig:slopes}.
The slope of the fundamental exhibits a minimum at some accretion rate and is strongly dependent on the geometrical parameters.
This behavior is a consequence of the scattering dip having a different energy dependence than the rest of the pulse profile.
The slope initially decreases because the depth of the dip depends on the energy dependent pulse amplitude, and begins to rise as the energy independent dip starts dominating the pulse profile.

\section{Discussion and conclusions}

Our simplified model for scattering in the accretion funnel has a significant effect on the pulse profile at accretion rates of $\gtrsim 10^{-10}~{M}_\sun\, \mathrm{yr}^{-1}$.
Accounting for this scattering in pulse profile fits could therefore lead to more accurate parameter estimates of AMPs at high accretion rates.
Performing model fits is beyond the scope of this paper, and should be a priority of future work.
The optical depth integral can feasibly be performed  quickly enough to be practical for Bayesian model fits, although it could be replaced with interpolation from a precalculated table.
The tables should be computed over a grid of $\Theta/\omega$ and $\alpha$ rather than $\Theta$ and $\psi_\infty$ to keep the parameters independent. 
The highly energy dependent nature of the dip may make it difficult to fit over large energy bins, because averaging pulse profiles over energy can smear the effect of the scattering.

Basing the model solely on the optical depth of scattering greatly simplifies the radiative transfer through the accretion funnel by not accounting for the scattered portion of the emission.
It correctly estimates the fraction of unscattered radiation and allows for simpler computations, but disregards the light that is scattered toward the observer.
The inclusion of light bending in the model would make solving the radiative transfer equation a much more formidable although not impossible problem.
While we account for the effect that the relativistic gas has on the opacity, in reality the influence it has on the angular distribution of radiation is more complex \citep{Beloborodov1998}.
The simplified model also cannot show what effects the scattering has on the polarization of the hot spot emission.
A more comprehensive radiative transfer model would be an important topic for future work, especially as phase-resolved polarization measurements are expected to improve the constraints on AMP geometrical parameters \citep{Salmi2021,Bobrikova2023}.

Modeling the accretion funnel as a finite cone allows   analytical computations  of the light trajectory to be performed, yet the true accretion geometry is more complex.
The real accretion flow curves toward the accretion disk following magnetic field lines, while a cone simply extends away from the hotspot.
The sharp maximum of the optical depth at low emission angles would not be present in a real funnel as it has no flat upper edge.
However, the largest scattering contribution happens near the NS surface so the geometry of the funnel near the inner disk radius matters less.
Thus, we can use the conical model to test the qualitative effects of the scattering optical depth, although we cannot be certain whether this accurately represents its dependence on the observer inclination and magnetic obliquity.
What consequence a different hotspot shape would have is also unclear under this approximation, and testing it would require a significantly more complex model.
The effect that the assumed hotspot and funnel geometry has on model fits should be investigated.

The observed soft time lags and pulse phases of AMPs show some dependence on the flux and, assuming that the flux is correlated with the accretion rate, we can compare this behavior with our model.
The observed timing noise of some AMPs also seems to be correlated with the X-ray flux during outbursts \citep{Patruno2010}, although testing this behavior with our model would be difficult.
A more easily testable example is the slope of the soft time lags of J1808, which decreases with flux up to a turnover point where the trend reverses.
This has been previously explained with the pulsar transitioning into the propeller regime, with the turnover caused by changes either in the accretion disk or the accretion funnel \citep{Hartman2009}.
Our model shows similar behavior when the accretion rate varies even with a static geometry (see Fig.~\ref{fig:slopes}).
The qualitative dependence of the slope on accretion rate would remain the same under a different accretion geometry, but it might act differently if the scattering was modeled as energy dependent.
The slope remains dependent on flux as long as the energy dependence of scattered light is different from the energy dependence of the incident light.

Our model does not predict any rapid phase shifts, and so they must be caused by other processes or by major changes in the geometry.
Some of the observed rapid shifts, such as that observed during the 2002 outburst of J1808, go in the opposite direction compared to the trend of the scattering model \citep{Ibragimov2009}.
We therefore conclude that whatever causes the rapid shifts often dominates the scattering.
The changes caused by scattering, although   weaker than other factors, should be relevant throughout the entire outburst of an AMP.

In conclusion, even our simplified approach to scattering in the accretion funnel demonstrates that it is an important factor in AMP pulse profiles.
A more accurate model for the polarized scattering would be of interest in anticipation of future observations.
It is also fascinating to see what changes it would have on NS parameter estimates.

\section*{Acknowledgments}

We acknowledge support from the Academy of Finland grant 333112. 
TS acknowledges support from ERC Consolidator Grant (CoG) No. 865768 AEONS (PI: A. Watts). 
Some of the theoretical work presented here was first done in the Master's thesis of \cite{Ahlberg2022}.
We would like to thank Iiro Vilja and Eugene Churazov for helpful advice.

\bibliographystyle{aa}
\bibliography{48153corr}

\begin{thebibliography}{40}
\expandafter\ifx\csname natexlab\endcsname\relax\def\natexlab#1{#1}\fi

\bibitem[{{Ahlberg}(2022)}]{Ahlberg2022}
{Ahlberg}, V. 2022, Master's thesis, University of Turku

\bibitem[{{AlGendy} \& {Morsink}(2014)}]{AlGendy2014}
{AlGendy}, M. \& {Morsink}, S.~M. 2014, \apj, 791, 78

\bibitem[{{Beloborodov}(1998)}]{Beloborodov1998}
{Beloborodov}, A.~M. 1998, \apjl, 496, L105

\bibitem[{Beloborodov(2002)}]{Beloborodov_2002}
Beloborodov, A.~M. 2002, \apj, 566, L85–L88

\bibitem[{{Bobrikova} {et~al.}(2023){Bobrikova}, {Loktev}, {Salmi}, \& {Poutanen}}]{Bobrikova2023}
{Bobrikova}, A., {Loktev}, V., {Salmi}, T., \& {Poutanen}, J. 2023, \aap, 678, A99

\bibitem[{{Chandrasekhar} \& {Breen}(1947)}]{Chandrasekhar1947}
{Chandrasekhar}, S. \& {Breen}, F.~H. 1947, \apj, 105, 435

\bibitem[{{Ghosh} \& {Lamb}(1978)}]{Ghosh1978}
{Ghosh}, P. \& {Lamb}, F.~K. 1978, \apjl, 223, L83

\bibitem[{{Gierli{\'n}ski} \& {Poutanen}(2005)}]{Gierlinski2005}
{Gierli{\'n}ski}, M. \& {Poutanen}, J. 2005, \mnras, 359, 1261

\bibitem[{{Hartman} {et~al.}(2008){Hartman}, {Patruno}, {Chakrabarty}, {Kaplan}, {Markwardt}, {Morgan}, {Ray}, {van der Klis}, \& {Wijnands}}]{Hartman2008}
{Hartman}, J.~M., {Patruno}, A., {Chakrabarty}, D., {et~al.} 2008, \apj, 675, 1468

\bibitem[{{Hartman} {et~al.}(2009{\natexlab{a}}){Hartman}, {Patruno}, {Chakrabarty}, {Markwardt}, {Morgan}, {van der Klis}, \& {Wijnands}}]{Hartman2009b}
{Hartman}, J.~M., {Patruno}, A., {Chakrabarty}, D., {et~al.} 2009{\natexlab{a}}, \apj, 702, 1673

\bibitem[{{Hartman} {et~al.}(2009{\natexlab{b}}){Hartman}, {Watts}, \& {Chakrabarty}}]{Hartman2009}
{Hartman}, J.~M., {Watts}, A.~L., \& {Chakrabarty}, D. 2009{\natexlab{b}}, \apj, 697, 2102

\bibitem[{Ibragimov \& Poutanen(2009)}]{Ibragimov2009}
Ibragimov, A. \& Poutanen, J. 2009, \mnras, 400, 492

\bibitem[{{Kajava} {et~al.}(2011){Kajava}, {Ibragimov}, {Annala}, {Patruno}, \& {Poutanen}}]{Kajava2011}
{Kajava}, J. J.~E., {Ibragimov}, A., {Annala}, M., {Patruno}, A., \& {Poutanen}, J. 2011, \mnras, 417, 1454

\bibitem[{{Kulkarni} \& {Romanova}(2005)}]{Kulkarni2005}
{Kulkarni}, A.~K. \& {Romanova}, M.~M. 2005, \apj, 633, 349

\bibitem[{{Kulkarni} \& {Romanova}(2013)}]{Kulkarni2013}
{Kulkarni}, A.~K. \& {Romanova}, M.~M. 2013, \mnras, 433, 3048

\bibitem[{{Lamb} {et~al.}(2009){Lamb}, {Boutloukos}, {Van Wassenhove}, {Chamberlain}, {Lo}, {Clare}, {Yu}, \& {Miller}}]{Lamb2009}
{Lamb}, F.~K., {Boutloukos}, S., {Van Wassenhove}, S., {et~al.} 2009, \apj, 706, 417

\bibitem[{{Leahy} {et~al.}(2008){Leahy}, {Morsink}, \& {Cadeau}}]{Leahy2008}
{Leahy}, D.~A., {Morsink}, S.~M., \& {Cadeau}, C. 2008, \apj, 672, 1119

\bibitem[{{Leahy} {et~al.}(2011){Leahy}, {Morsink}, \& {Chou}}]{Leahy2011}
{Leahy}, D.~A., {Morsink}, S.~M., \& {Chou}, Y. 2011, \apj, 742, 17

\bibitem[{{Loktev} {et~al.}(2020){Loktev}, {Salmi}, {N{\"a}ttil{\"a}}, \& {Poutanen}}]{Loktev2020}
{Loktev}, V., {Salmi}, T., {N{\"a}ttil{\"a}}, J., \& {Poutanen}, J. 2020, \aap, 643, A84

\bibitem[{{Long} {et~al.}(2012){Long}, {Romanova}, \& {Lamb}}]{Long2012}
{Long}, M., {Romanova}, M.~M., \& {Lamb}, F.~K. 2012, \na, 17, 232

\bibitem[{{Lyubarskii} \& {Syunyaev}(1982)}]{Lyubarskiy82}
{Lyubarskii}, Y.~E. \& {Syunyaev}, R.~A. 1982, Soviet Astronomy Letters, 8, 330

\bibitem[{{Misner} {et~al.}(1973){Misner}, {Thorne}, \& {Wheeler}}]{Misner73}
{Misner}, C.~W., {Thorne}, K.~S., \& {Wheeler}, J.~A. 1973, {Gravitation} (San Francisco: W.H.~Freeman and Co.)

\bibitem[{{Mushtukov} {et~al.}(2015){Mushtukov}, {Tsygankov}, {Serber}, {Suleimanov}, \& {Poutanen}}]{Mushtukov15}
{Mushtukov}, A.~A., {Tsygankov}, S.~S., {Serber}, A.~V., {Suleimanov}, V.~F., \& {Poutanen}, J. 2015, \mnras, 454, 2714

\bibitem[{{N{\"a}ttil{\"a}} {et~al.}(2017){N{\"a}ttil{\"a}}, {Miller}, {Steiner}, {Kajava}, {Suleimanov}, \& {Poutanen}}]{Nattila2017}
{N{\"a}ttil{\"a}}, J., {Miller}, M.~C., {Steiner}, A.~W., {et~al.} 2017, \aap, 608, A31

\bibitem[{{Patruno} {et~al.}(2010){Patruno}, {Hartman}, {Wijnands}, {Chakrabarty}, \& {van der Klis}}]{Patruno2010}
{Patruno}, A., {Hartman}, J.~M., {Wijnands}, R., {Chakrabarty}, D., \& {van der Klis}, M. 2010, \apj, 717, 1253

\bibitem[{{Patruno} \& {Watts}(2021)}]{Patruno2021}
{Patruno}, A. \& {Watts}, A.~L. 2021, in Astrophysics and Space Science Library, Vol. 461, Timing Neutron Stars: Pulsations, Oscillations and Explosions, ed. T.~M. {Belloni}, M.~{M{\'e}ndez}, \& C.~{Zhang}, 143--208

\bibitem[{{Pechenick} {et~al.}(1983){Pechenick}, {Ftaclas}, \& {Cohen}}]{Pechenick83}
{Pechenick}, K.~R., {Ftaclas}, C., \& {Cohen}, J.~M. 1983, \apj, 274, 846

\bibitem[{{Poutanen}(2020)}]{Poutanen2020}
{Poutanen}, J. 2020, \aap, 640, A24

\bibitem[{{Poutanen} \& {Beloborodov}(2006)}]{Poutanen2006}
{Poutanen}, J. \& {Beloborodov}, A.~M. 2006, \mnras, 373, 836

\bibitem[{{Poutanen} \& {Gierli{\'n}ski}(2003)}]{Poutanen2003}
{Poutanen}, J. \& {Gierli{\'n}ski}, M. 2003, \mnras, 343, 1301

\bibitem[{{Poutanen} {et~al.}(2009){Poutanen}, {Ibragimov}, \& {Annala}}]{Poutanen2009}
{Poutanen}, J., {Ibragimov}, A., \& {Annala}, M. 2009, \apjl, 706, L129

\bibitem[{{Romanova} {et~al.}(2004){Romanova}, {Ustyugova}, {Koldoba}, \& {Lovelace}}]{Romanova2004}
{Romanova}, M.~M., {Ustyugova}, G.~V., {Koldoba}, A.~V., \& {Lovelace}, R.~V.~E. 2004, \apj, 610, 920

\bibitem[{{Romanova} {et~al.}(2003){Romanova}, {Ustyugova}, {Koldoba}, {Wick}, \& {Lovelace}}]{Romanova2003}
{Romanova}, M.~M., {Ustyugova}, G.~V., {Koldoba}, A.~V., {Wick}, J.~V., \& {Lovelace}, R.~V.~E. 2003, \apj, 595, 1009

\bibitem[{{Salmi} {et~al.}(2021){Salmi}, {Loktev}, {Korsman}, \& et. al.}]{Salmi2021}
{Salmi}, T., {Loktev}, V., {Korsman}, K., \& et. al. 2021, \aap, 646, A23

\bibitem[{{Suleimanov} {et~al.}(2011){Suleimanov}, {Poutanen}, {Revnivtsev}, \& {Werner}}]{Suleimanov2011}
{Suleimanov}, V., {Poutanen}, J., {Revnivtsev}, M., \& {Werner}, K. 2011, \apj, 742, 122

\bibitem[{{Suleimanov} {et~al.}(2017){Suleimanov}, {Poutanen}, {N{\"a}ttil{\"a}}, {Kajava}, {Revnivtsev}, \& {Werner}}]{Suleimanov2017}
{Suleimanov}, V.~F., {Poutanen}, J., {N{\"a}ttil{\"a}}, J., {et~al.} 2017, \mnras, 466, 906

\bibitem[{{Suleimanov} {et~al.}(2020){Suleimanov}, {Poutanen}, \& {Werner}}]{Suleimanov2020}
{Suleimanov}, V.~F., {Poutanen}, J., \& {Werner}, K. 2020, \aap, 639, A33

\bibitem[{{Viironen} \& {Poutanen}(2004)}]{Viironen2004}
{Viironen}, K. \& {Poutanen}, J. 2004, \aap, 426, 985

\bibitem[{{Watts} {et~al.}(2016){Watts}, {Andersson}, {Chakrabarty}, {Feroci}, {Hebeler}, {Israel}, {Lamb}, {Miller}, {Morsink}, {{\"O}zel}, {Patruno}, {Poutanen}, {Psaltis}, {Schwenk}, {Steiner}, {Stella}, {Tolos}, \& {van der Klis}}]{Watts2016}
{Watts}, A.~L., {Andersson}, N., {Chakrabarty}, D., {et~al.} 2016, Reviews of Modern Physics, 88, 021001

\bibitem[{{Watts} {et~al.}(2019){Watts}, {Yu}, {Poutanen}, {Zhang}, {Bhattacharyya}, {Bogdanov}, {Ji}, {Patruno}, {Riley}, {Bakala}, {Baykal}, {Bernardini}, {Bombaci}, {Brown}, {Cavecchi}, {Chakrabarty}, {Chenevez}, {Degenaar}, {Del Santo}, {Di Salvo}, {Doroshenko}, {Falanga}, {Ferdman}, {Feroci}, {Gambino}, {Ge}, {Greif}, {Guillot}, {Gungor}, {Hartmann}, {Hebeler}, {Heger}, {Homan}, {Iaria}, {Zand}, {Kargaltsev}, {Kurkela}, {Lai}, {Li}, {Li}, {Li}, {Linares}, {Lu}, {Mahmoodifar}, {M{\'e}ndez}, {Coleman Miller}, {Morsink}, {N{\"a}ttil{\"a}}, {Possenti}, {Prescod-Weinstein}, {Qu}, {Riggio}, {Salmi}, {Sanna}, {Santangelo}, {Schatz}, {Schwenk}, {Song}, {{\v{S}}r{\'a}mkov{\'a}}, {Stappers}, {Stiele}, {Strohmayer}, {Tews}, {Tolos}, {T{\"o}r{\"o}k}, {Tsang}, {Urbanec}, {Vacchi}, {Xu}, {Xu}, {Zane}, {Zhang}, {Zhang}, {Zhang}, {Zheng}, \& {Zhou}}]{Watts2019}
{Watts}, A.~L., {Yu}, W., {Poutanen}, J., {et~al.} 2019, Science China Physics, Mechanics, and Astronomy, 62, 29503

\end{thebibliography}

\end{document}